\documentclass[11pt]{article}

\usepackage{euscript}
\usepackage{amssymb}
\usepackage{amsfonts}
\usepackage{amsbsy}
\usepackage{amsmath}
\usepackage{epsfig}
\usepackage{amsthm}
\usepackage{amscd}
\usepackage{amstext}
\usepackage{verbatim}

 \usepackage{tensor}
 \usepackage{color}
 \usepackage{graphicx}
 \usepackage{caption}
 \usepackage{subcaption}
 \usepackage{stmaryrd}

\usepackage[pdftex]{hyperref}

\def\ben{\begin{equation}}
\def\een{\end{equation}}

 \let\h=\eta  
    \let\p=\phi

\let \h=\hat
\def\be{\begin{equation}}
\def\ee{\end{equation}}
\def\beq{\begin{equation}}
\def\eeq{\end{equation}}
\def\ba{\begin{array}}
\def\ea{\end{array}}

\def\dalemb#1#2{{\vbox{\hrule height .#2pt
       \hbox{\vrule width.#2pt height#1pt \kern#1pt
               \vrule width.#2pt}
       \hrule height.#2pt}}}

\newcommand{\bea}{\begin{eqnarray}}
\newcommand{\eea}{\end{eqnarray}}

\renewcommand{\p}{\partial}

\newcommand{\tP}{\tilde{P}}
\newcommand{\tL}{\tilde{L}}

\newcommand{\refe}[1]{(\ref{#1})}

\newcommand{\ra}{\rightarrow}

\begin{document}

\begin{center}

{ \LARGE {\bf Phase Transitions in Warped AdS$_3$ Gravity}}

\vspace{1.2cm}

St\'{e}phane Detournay$^\sharp$ and C\'eline Zwikel$^\sharp$

\vspace{0.9cm}

{\it $^\sharp$ Physique Th\'eorique et Math\'ematique\\
Universit\'e Libre de Bruxelles and International Solvay Institutes \\Campus Plaine C.P.~231, B-1050 Bruxelles, Belgium}






\vspace{1.6cm}

{\tt sdetourn, czwikel@ulb.ac.be} \\

\vspace{1.6cm}

\end{center}

\begin{abstract}
We consider asymptotically Warped AdS$_3$ black holes in Topologically Massive Gravity. We study their thermodynamic stability and show the existence of a Hawking-Page phase transition between the black hole and thermal background phases. At zero angular potential, the latter is shown to occur at the self-dual point of the dual Warped Conformal Field Theory partition function, in analogy with the phase transition for BTZ black holes in AdS$_3$/CFT$_2$. We also discuss how the central and vacuum charges  can be obtained from inner horizon mechanics in the presence of a gravitational anomaly.

\end{abstract}
\pagebreak
\setcounter{page}{1}

\tableofcontents

\newpage
\section{Introduction}

The occurrence of phase transitions in gravitational systems has been known since the seminal work of Hawking and Page more than 30 years ago \cite{hawking1982}. They uncovered a thermal phase transition in asymptotically anti-de Sitter (AdS) spaces: when thermal radiation in AdS is heated up beyond a certain temperature, it becomes unstable and eventually collapses to form a black hole. This statement has been generalized to various setups 
and entails many interesting consequences. For instance, in the context of AdS/CFT \cite{Maldacena:1997re, Witten:1998qj, Gubser:1998bc} where quantum gravity in AdS space is conjectured to be dual to a Conformal Field Theory (CFT), the Hawking-Page transition is interpreted as a confining/deconfining phase transition in the dual field theory at high temperature \cite{Witten:1998zw}. 
The Hawking-Page phase transition in AdS$_3$ gravity (between BTZ black holes \cite{Banados:1992wn, Banados:1992gq} and thermal AdS$_3$) has also been studied from various perspectives and shown to exhibit a surprisingly rich structure \cite{Maldacena:1998bw, Dijkgraaf:2000fq, Maloney:2007ud} (see also \cite{Berkooz:2007fe, Kleban:2004rx}). Through AdS$_3$/CFT$_2$, the AdS$_3$ quantum gravity partition function should be equal to a CFT$_2$ partition function of the form
\be \label{cftpartitionfunction}
 Z(\tau, \bar \tau) = \mbox{Tr} e^{2 \pi i \tau L_0} e^{2 \pi i \bar \tau \bar L_0},
 \ee
  where $\tau$ is the modular parameter of the torus on which the 2d CFT is defined. On the gravity side the grand canonical partition function is usually expressed, for want of anything better, as a euclidean path integral $Z(\tau) \sim \int {\cal D}g e^{-k S[g]}$ over all (smooth) metrics with boundary metric a two torus with conformal structure $\tau$. The latter encodes the thermal properties of the geometries under consideration through the identifications $z \sim z + 1 \sim z + \tau$ with $z = \frac{1}{2 \pi} (\phi + i t)$ and $\tau = \frac{1}{2 \pi} (\Theta + i \beta)$, where $\beta$ is the dimensionless inverse temperature and $\Theta$ the angular potential. In the classical, large AdS-radius, large central charge limit the partition function can be approximated in the saddle-point approximation as
\be \label{adspartitionfunction}
   Z(\tau, \bar \tau) \sim \sum_{g_c} e^{-k S[g_c]}, \quad k = \frac{c}{24} = \frac{\ell}{16 G} \gg 0,
\ee
where $c$ is the Brown-Henneaux central charge \cite{Brown:1986nw} and $S[g]$ the classical gravity action. This is the regime in which we will be working. The Gibbs free energy $G(\tau, \bar \tau)$ of the system is defined as
\be \label{freeen}
  Z(\tau, \bar \tau) = e^{-\beta G(\tau, \bar \tau)}.
\ee
A crucial property of 2d CFTs whose partition functions can be obtained as the vacuum amplitude on the torus is \emph{modular invariance}. In particular, the modular S-transformation implies that 
\be \label{modularinvariance}
  Z(\tau, \bar \tau) =  Z(-\frac{1}{\tau},- \frac{1}{\bar \tau}).
\ee
At zero angular potential $\Theta$, this powerful property relates the behaviour of the partition function at high and low temperatures. In these regimes, the free energy for \emph{generic} CFTs is universal and depends only on the central charges. Interestingly, it was shown \cite{Keller:2011xi} (see also \cite{Hartman:2014oaa}) that for \emph{holographic} CFTs (those dual to AdS$_3$ gravity backgrounds), in the large central charge limit, this universal behaviour extends all the way towards the self-dual temperature
\be \label{sdbeta}
  \beta_{sd} = 2 \pi.
\ee
This peculiar behaviour of the free energy allows to explain why, while generic 2d CFTs have no phase transition ($\ln Z$ and all its derivatives are continuous functions of $\beta$), holographic CFTs do exhibit a first order phase transition at \refe{sdbeta} at large $c$. This regime is precisely the one of classical gravity and \refe{sdbeta} then precisely corresponds to the AdS$_3$ Hawking-Page transition \cite{Keller:2011xi}  where the classical actions of the black hole and thermal AdS$_3$ saddles coincide.


Modular invariance \refe{modularinvariance} is also at the heart of another striking result pertaining to unitary 2d CFTs, the derivation of the asymptotic growth of states through Cardy's formula \cite{Cardy:1986ie}. For large $L_0$ and $\bar L_0$ charges and fixed central charges, the result is simply
\be \label{cardy}
   S_{CFT}(L_0, \bar L_0) = 2 \pi \sqrt{\frac{c}{6} L_0} + 2 \pi \sqrt{\frac{\bar c}{6} \bar L_0}.
\ee
This formula has deep implications for the black hole entropy problem. Indeed, while in some situations the dual CFT can be thoroughly identified (the most famous example being the symmetric orbifold CFT in the microscopic derivation of \cite{Strominger:1996sh}), in many cases it can not. However even then, expression \refe{cardy} was shown to match the Bekenstein-Hawking of AdS$_3$ black holes, based only upon symmetry arguments \cite{Strominger:1997eq}. Moreover, it does so \emph{for all} values of the charges. The reason was elucidated recently in details \cite{Hartman:2014oaa}: in holographic CFTs  (defined more precisely as admitting a large central charge expansion and a sparse light spectrum), the Cardy formula was shown to hold in the large $c$ limit for $L_0$, $\bar L_0$ of order $c$, which is precisely the regime in which the Bekenstein-Hawking entropy is computed. 

The upshot of the above discussion is that gravity computations strongly constrain the properties that a holographic field theory dual to gravity in asymptotically AdS spaces should display. On the other hand, in recent years, the relevance of gravity theories with non-AdS boundary conditions have appeared in various holographic contexts. Examples include dS/CFT \cite{Strominger:2001pn}, Kerr/CFT \cite{Guica:2008mu}, AdS/Condensed Matter Theory \cite{Son:2008ye, Balasubramanian:2008dm} and more recently flat space Holography (see e.g.\cite{Barnich:2012xq, Bagchi:2012xr} and references therein). In most of these situations, not much is known about the potential dual field theories which are not expected to be (at least full-fledged) CFTs. In this work, we will focus on a class of spaces called \emph{Warped AdS$_3$} spaces (WAdS$_3$). They have appeared at various occasions, namely in string theory \cite{Israel:2004vv, Israel:2003cx, Compere:2008cw, Detournay:2010rh, Azeyanagi:2012zd, Detournay:2012dz} and lower-dimensional gravity \cite{Rooman:1998xf, Moussa:2003fc, Banados:2005da, Compere:2007in, Bouchareb:2007yx, Anninos:2008fx, Compere:2009zj, Chen:2013aza}. They are believed to be relevant both to describe extremal black holes holographically as in Kerr/CFT and in Holographic Condensed Matter applications where they appear as factors of the corresponding geometries \cite{ElShowk:2011cm, Detournay:2012pc, Detournay:2012dz}. As alluded to above, 
they do not satisfy Brown-Henneaux's boundary conditions, but rather belong to a phase space with $Vir \inplus \hat u(1) $ symmetry \cite{Compere:2007in, Compere:2008cv, Compere:2009zj}. These symmetries were shown to follow from chiral scale invariance in 2d field theories \cite{Hofman:2011zj}, in a way similar to the emergence of full local conformal symmetry in 2d Poincar\' e and scale invariant field theories. Two-dimensional field theories with $Vir \inplus \hat u(1)$ local symmetries have been introduced in \cite{Detournay:2012pc} and named \emph{Warped Conformal Field Theories} (WCFTs). They share many remarkable features with 2d CFTs, as we will review in more details in the next section. In particular, they exhibit a counterpart of the modular invariance \refe{modularinvariance} and the symmetries are powerful enough to derive an analog of Cardy's formula \cite{Cardy:1986ie}, denoted $S_{WCFT}$. Furthermore, the Bekenstein-Hawking entropy of the black holes contained in the gravitational phase space is exactly equal to $S_{WCFT}$.

The goal of this paper is to investigate the existence of phase transitions in WAdS$_3$ gravity and relate them to properties of the putative underlying WCFT. It is fair to say that even generic WCFTs are rather elusive and explicit examples are scarce (see however \cite{Compere:2013aya, Hofman:2014loa}). We will therefore use the gravity side as a guide to explore the basic requirements a holographic WCFT should satisfy. As WAdS$_3$ spaces do not satisfy Einstein's equations, we will use the simplest, purely metric, gravitational theory admitting them as solutions: Topologically Massive Gravity (TMG) \cite{Deser:1981wh, Deser:1982vy}. The latter consists in a certain higher-curvature, parity breaking, extension of Einstein gravity, which we will review in due course. 

The plan of the paper is as follows. In Sect.~\ref{sectthermal}, we review some salient properties of WCFTs, discussing in particular the existence of inequivalent thermal ensembles and the counterpart of modular invariance. Sect.~\ref{btztmg} introduces the gravity theory under consideration, considering as a warm-up the thermodynamic stability properties of BTZ black holes and the Hawking-Page phase transition for AdS$_3$ in TMG. With the technology introduced in previous section, Sect.~\ref{wbtztmg} finally adresses the same questions for WAdS$_3$ black holes and connects the results to properties of the WCFT partition function. In Sect.~\ref{SectInner} we make a brief digression on inner horizon properties, and then conclude.

\section{Warped AdS$_3$ Black Holes and WCFT Thermal Ensembles}\label{sectthermal}


WAdS$_3$ backgrounds are Lorentzian counterparts of the squashed three-sphere, and are obtained by deforming the AdS$_3$ metric as follows:
\be \label{wads3}
  g_{WAdS} = g_{AdS_3} - 2 H^2 \xi \otimes \xi,
\ee
where the real constant $H^2$ is a deformation parameter and where $\xi^\mu$ is a constant-norm Killing vector, belonging to one $SL(2,\mathbb{R})$ factor of the $SL(2,\mathbb{R}) \times SL(2,\mathbb{R})$ isometry group of AdS$_3$. By construction, the resulting geometry possesses an $SL(2,\mathbb{R}) \times U(1)$ isometry group, and is named Timelike, Spacelike or Null Warped AdS$_3$ depending on the norm of the deformation Killing vector ($||\xi||^2= -1, +1$ or $0$ respectively). These metrics no longer solve Einstein's equations, but are a solution to the equations of motion of various higher-curvature extensions of 3d gravity and of Einstein-Hilbert gravity coupled to matter fields.

Locally WAdS$_3$ black holes can be obtained by quotienting the above (spacelike and null) warped geometries, much like the BTZ black holes correspond to discrete quotients of AdS$_3$ \cite{Banados:1992wn,Banados:1992gq}. Here we will focus on the spacelike case. Various coordinate systems have been used in the literature to describe these black holes.
Their metric can be written as (see \cite{Anninos:2008fx}, Sect. 4)
\begin{align}  \nonumber
\frac{ds^2}{\h l^2}=
dT^2 & +\frac{d\hat {r}^2}{(\nu^2+3)(\hat r-\hat r_+)(\hat r-\hat r_-)}-\left( 2\nu\hat r -\sqrt{\hat r_+ \hat r_-(\nu^2+3)}\right) dT d\theta \\\label{0807.3040BH}
&+\frac {\hat r} 4 \left( 3(\nu^2-1)\hat r+(\nu^2+3)(\hat r_++\h r_-)-4\nu\sqrt{\hat r_+ \hat r_-(\nu^2+3)  } \right)
\end{align}
with $\hat{r} \in [0, \infty[$, $T \in ]-\infty, +\infty[$ and $\theta \sim \theta + 2 \pi$. The solution is free of naked CTCs when the parameter $\nu^2 > 1$, in which case the solution is said to be spacelike \emph{stretched} (otherwise it is said to be  \emph{squashed} -- this terminology originates from the fact that the warp factor deforming the AdS$_3$ can be greater or less than unity, see \cite{Anninos:2008fx}; spacelike squashed black holes are sometimes referred to as Godel black holes \cite{Banados:2005da, Compere:2008cw}). The parameter $\nu$ is in general fixed by the equations of motion in terms of the coupling constants of the theory at hand and characterizes the deformation like the parameter $H^2$ in (\ref{wads3}). In particular, $\nu^2 = 1$ when $H^2=0$ and we recover a locally AdS$_3$ space (\eqref{0807.3040BH} then represents a BTZ black hole in unusual coordinates). 

The solutions \refe{0807.3040BH} have well-defined thermodynamic properties. They are characterized by a Hawking temperature, angular velocity, mass, angular momentum and Bekenstein-Hawking entropy satisfying the first law of black hole mechanics. Their asymptotic behaviour, however, is unusual as can be seen from the leading terms in $\hat{r}$ in \refe{0807.3040BH}: when $\nu^2 \ne 1$, they cease to be asymptotically AdS$_3$, i.e. they no longer satisfy Brown-Henneaux's boundary conditions \cite{Brown:1986nw}. Instead, they belong to the phase space of asymptotically spacelike WAdS$_3$ spaces (see (1.1) of \cite{Compere:2009zj}). These boundary conditions are preserved by the following set of diffeomorphisms
\bea \label{AKV}
\ell_n &=& e^{i n \theta}\p_\theta  - i n \hat{r} e^{i n \theta}\p_{\hat{r}} \\
p_n &=& e^{i n \theta}\p_T  \ .\notag
\eea
generating a centerless Virasoro-Kac-Moody $U(1)$ algebra
\be \label{viru1}
i [\ell_n\, \ell_m ] = (n-m)\ell_{n+m} \ , \quad\quad i [\ell_n,  p_m]= -m \ p_{n+m} \ ,  \quad\quad i [p_n,  p_m]=0 \ .
\ee
The corresponding canonical charges 
\be \label{wcharges}
L_n := Q_{\ell_n} \ , \quad \quad P_n := Q_{p_n} \ ,
\ee
satisfy the same algebra under Dirac brackets, with central extensions which we denote by $c$ and $k$ for the Virasoro and $U(1)$ Kac-Moody parts respectively. 
Their precise expressions depend on the details of the gravitational theory under consideration, see e.g. \cite{Compere:2008cv, Compere:2009zj, Detournay:2012pc, Detournay:2012dz}.

Another useful description of the black holes \refe{0807.3040BH} is by writing 
\be \label{WBTZ}
   ds^2_{WBTZ} = ds^2_{BTZ} - 2 H^2 \xi \otimes \xi
\ee
where $ds^2_{BTZ}$ is the BTZ black hole metric (where we put Newton's constant $G$ to 1):
\be \label{BTZ}
ds^2_{BTZ} = \left(8M-\frac{r^2}{l^2}\right)  dt^2-\frac{ r^2 dr^2}{8  M r^2-\frac{r^4}{l^2}-16 J^2 } +8 J\,  dt\,  d\phi+r^2 d\phi^2
\ee
and $\xi$ such that $||\xi||^2 = 1$ is given by
\be \label{xibtz}
\xi^\mu=\frac1{\sqrt8}\sqrt{\frac{ l}{(M \,  l -J)} } \left(-\partial_t+\partial_\phi\right). 
\ee
This description makes the relation to BTZ and asymptotically AdS$_3$ spaces more direct. The change of boundary coordinates between \refe{0807.3040BH} and \refe{WBTZ} is (see Appendix 1 for details)
\bea \label{changecoords}
  \phi - \frac{t}{l} &=&-\h l \, \theta \\
  \phi + \frac{t}{l} &=& -\h l \left( \frac{k}{2 P_0} T + \theta \right) \nonumber
\eea
with $l=\sqrt{\frac4{3+\nu^2}} \hat l$. We will often also set $l=1$.

This change of coordinates is charge-dependent and will affect the form of the corresponding boundary conditions, which take the form  ($x^\pm = \phi \pm \frac{t}{l}$)
\bea
g_{rr} &=& \frac{l^2}{r^2} + O(r^{-4}), \; \;g_{++} =  j_{++} r^4 + h_{++} r^2 + f_{++}, \;\;\; g_{+-} = (\frac{1}{2} + j_{+-}) r^2 + O(1), \nonumber\\  g_{--} &=& f_{--}, \; \;\;g_{+r} = O(1/r),\;\; \;g_{-r} = O(1/r^3). \label{FullBC}
\eea


The change of coordinates \refe{changecoords} together with \refe{wcharges} implies that \cite{Detournay:2012pc}
\be \label{xpxm}
   \tilde{P_0} := Q_{\p_{x^+}} = \frac{P_0^2}{k}, \qquad \tilde{L_0} := Q_{\p_{x^-}} = L_0 - \frac{P_0^2}{k}.
\ee




In usual AdS/CFT set-ups, certain CFT density matrices on the field theory side correspond to specific holographically dual geometries (more specifically, the latter actually arise as saddle points in a quantum gravity path integral). 
In three dimensions for instance, the CFT$_2$ vacua on the plane or on the cylinder ($\rho = |0\rangle\langle0|$), thermal density matrices ($\rho = e^{-\beta H}$) and grand canonical density matrices ($\rho = e^{-\beta (H - \Omega J)}$) are dual to Poincar\'e or global AdS$_3$, static BTZ and rotating BTZ geometries, respectively. On the other hand multi-particle descendant states of the form $|\psi \rangle = L_{-n_1}...L_{-n_m}|0\rangle$ are dual to AdS$_3$ dressed up with boundary gravitons, i.e. AdS$_3$ acted upon with finite Brown-Henneaux diffeomorphisms. 
The field theory is understood to ``live" at the conformal boundary of AdS$_3$, whose coordinates are identified with the field theory coordinates. In particular, the change of coordinates between BTZ black holes and Poincar\'e AdS$_3$, close to the boundary, is precisely the one between Rindler and Minkowski observers in (1+1)-dimensions \cite{Maldacena:1998bw}.  
Therefore, having a BTZ black hole in the bulk corresponds to having a field theory at the boundary as seen by an accelerated observer and the field theory will be in a thermal state.

The above statement relies essentially on the relation between Rindler and inertial observers and the existence of a UV dual field theory at large $r$, but not on its exact nature (conformal or not). A similar argument can be performed for WAdS$_3$ spaces \cite{Detournay:2012pc}. 
First, the symmetries \refe{AKV} suggest that asymptotically spacelike WAdS$_3$ spaces are dual to a field theory with the corresponding symmetries, a Warped Conformal Field Theory (WCFT) \cite{Detournay:2012pc}. By analogy with the AdS$_3$ situation, its vacuum on the plane is taken to be dual to Poincar\'e spacelike WAdS$_3$, with metric
\be\label{pwads3}
   ds^2_{PWAdS_3} = \frac{dy^2 + dw^+ dw^-}{y^2} - 2 H^2 \xi \otimes \xi, 
\ee
with
\be
 \xi^\mu = y \p_y + 2 w^- \p_-, \qquad ||\xi||^2 = 1.
\ee
Second, the change of coordinates between \refe{pwads3} and \refe{WBTZ} is 
\bea \label{wrindler}
  w^+ &=&\sqrt{\frac{r^2-l^4\pi(T_++T_-)^2}{r^2-l^4\pi(T_+-T_-)^2}} \,e^{2 \pi l \,T_+ x^+} \nonumber\\
  w^-  &=&\sqrt{\frac{r^2-l^4\pi(T_++T_-)^2}{r^2-l^4\pi(T_+-T_-)^2}} \, e^{2 \pi l \,T_- x^-} \nonumber\\ 
  y  &=&2\pi^2l^2\sqrt{\frac{T_-T_+}{r^2-l^4\pi(T_+-T_-)^2}}\, e^{\pi l\,(T_-x_-+T_+x_+)} 
\eea
where the parameters $T_{\pm}$ are given by
\be
 T_+=\frac T{1-l\, \Omega}, \quad T_-=\frac T{1+l\, \Omega}.
\ee
Near the $r=\infty$ boundary, \refe{wrindler} is just the change of coordinates from Minkowski to Rindler space\footnote{Rindler space usually depends on one parameter, the Rindler temperature (or acceleration). This case corresponds to $T_+ = T_-$ in the above and yields a thermal density matrix $\rho = e^{-\beta H}$ for the Minkowski vacuum as seen by the Rindler observer. When $T_+ \ne T_-$, the resulting state is $\rho = e^{-\beta (H - \Omega J)}$, with $T^{-1} =: \beta$ and $\Omega$ given by \refe{potentials}.}, so an observer in $(x^+, x^-)$ coordinates will be in a thermal state described by the density matrix $\rho = e^{-\beta_+ \tilde{P_0} - \beta_-  \tilde{L_0} }$, with 
\be
  \beta_{\pm} = \frac{1}{l\, T_\pm}.
\ee
Because of \refe{xpxm}, this explains why the coordinate system of \refe{WBTZ} is better suited to describe the thermal properties of the Warped black holes.

Note that the thermal identifications $(t,\phi) \sim (t + i \beta, \phi - i \beta \Omega)$, where $\beta$ and $\Omega$ are the inverse Hawking temperature and angular potential of \refe{WBTZ} respectively, are trivial in the $(w^+, w^-)$ plane because of \refe{wrindler} and the relations 
\be \label{potentials}
   \frac{1}{T} = \frac12\left (\frac{1}{T_-} + \frac{1}{T_+}\right ), \quad   \frac{\Omega \, l}{T} =\frac12\left ( \frac{1}{T_-} - \frac{1}{T_+}\right )
 \ee
%


Finally, the partition function for a WCFT at finite temperature and angular potential can be written as
\be \label{quadens}
   Z(\beta_+,\beta_-) = \mbox{Tr} e^{-\beta_+ \tilde{P_0} - \beta_-  \tilde{L_0}}.
\ee
The WCFT symmetries allow to determine the counterpart of modular invariance in these theories as \cite{Detournay:2012pc}
\be \label{wmodularquad}
Z(\beta_+,\beta_-) = Z\left (\frac{4 \pi^2}{\beta_+},\frac{4 \pi^2}{\beta_-}\right ) \,.
\ee
Defining $\tau = \frac{1}{2 \pi} (-\beta \Omega_E + i \frac{\beta}{l})$, with $\Omega_E = i \Omega$ the Wick-rotated angular potential, \refe{wmodularquad} is rewritten using \refe{potentials} as
\be
  Z(\tau, \bar \tau) = Z (-\frac{1}{\tau}, -\frac{1}{\bar \tau})
\ee
and therefore takes exactly the same form as in 2d CFTs. In particular, at zero angular potential, $\beta_+ = \beta_- = \beta$ and \refe{wmodularquad} singles out a self-dual temperature
\be \label{wsdtemp}
  \beta_{wsd} = 2 \pi,
\ee
the same way as in 2d CFTs.

The warped modular invariance relation then allows to obtain an expression for the degeneracy of states as 
\be \label{wcftentropy}
S_{WCFT}= 4\pi \sqrt{-\tP_0^{vac} \tP_0} + 4\pi \sqrt{-\tL_0^{vac} \tL_0} \,. 
\ee 
In deriving \refe{wcftentropy}, a \emph{small angular potential limit} has been taken (similar to the high-temperature limit of Cardy's formula) and the operators $\tilde{L_0}$ and $\tilde{P_0}$ were assumed to be bounded from below (which in gravity language coincides with the requirement that the metrics \eqref{0807.3040BH} display a horizon hiding the singularity), taking the values $\tilde{P_0^{vac}}$ and $\tilde{L_0^{vac}}$ on the ground state.

Defining quantities $c_R$ and $c_L$ through
\be \label{warpedclcr}
  \tilde{L_0^{vac}} = -\frac{c_R}{24}, \quad  \tilde{P_0^{vac}} = -\frac{c_L}{24},
\ee
the resulting expression \refe{wcftentropy} looks exactly like the Cardy formula for a 2d CFT. However, its origin is different since it derives from the WCFT symmetries, which are purely right-moving. Furthermore, in unitary 2d CFTs, the vacuum values of the Virasoro zero modes are univoquely determined in terms of the central charges. The situation is different in WCFTs: there, as discussed in \cite{Detournay:2012pc} $\tilde{L_0^{vac}}$ is fixed in terms of the Virasoro central extension $c$, while $\tilde{P_0^{vac}}$ depends on the details of the theory and is not determined by symmetries alone. In Holographic WCFTs, $\tilde{P_0^{vac}}$ can be computed from the gravitational charges of the vacuum geometry. The latter corresponds to \refe{WBTZ} for $M=-1/(8\, l)$ and $J=0$. At those values the resulting geometry has an enhanced $SL(2,R) \times U(1)$ isometry and is smooth. Then the Bekenstein-Hawking entropy of the black holes \refe{0807.3040BH} is found to match the WCFT one \cite{Detournay:2012pc}:
\be\label{BHWCFT}
   S_{BH} = S_{WCFT}.
\ee
The validity of \refe{BHWCFT} actually extends beyond the slowly-rotating regime, the same way the Cardy formula's validity extends beyond the high-temperature limit to explain the Bekenstein-Hawking entropy of AdS$_3$ black holes. Similarly to the AdS$_3$ case \cite{Hartman:2014oaa}, this is likely to constrain the spectrum of Holographic WCFTs.

In the following, we will be interested in studying possible phase transitions in holographic WCFTs between Warped black holes and the global vacuum put at finite potentials.


%




\section{BTZ Black Holes in TMG} \label{btztmg}
In this section, we analyse the thermodynamics and Hawking-Page transition of the BTZ black hole in Topologically Massive Gravity (TMG). We first review some relevant aspects of TMG and then present the thermodynamic quantities for the BTZ back hole and AdS$_3$ in TMG. These quantities will be used to analyse the local stability of BTZ and also the Hawking-Page transition. 

The TMG action is given by the Einstein-Hilbert action plus a gravitational Chern-Simons term 
\begin{equation}
  S=\frac1{16\pi}  \int d^3x\sqrt{-g}(R-2\Lambda)
  + \frac1{16\pi }\frac1{2\mu}  \int d^3x\sqrt{-g}\epsilon^{\lambda\mu\nu}\Gamma^\rho_{\lambda\sigma}\left(\partial_\mu\Gamma^\sigma _{\rho\nu}+ \frac23\Gamma^\sigma _{\mu\tau}\Gamma^\tau_{\nu\rho} \right)
\end{equation}
with $\mu$ the Chern-Simons coupling who can be taken positive without loss of generality. It is important to notice that the Chern-Simons term breaks parity. Pure general relativity (GR) is recovered in the limit $\mu\rightarrow \infty$.
The equations of motion are 
  \begin{equation}
  R_{\mu\nu} - \frac12 g_{\mu\nu}R + \Lambda g_{\mu\nu} +\frac1\mu C_{\mu\nu}=0
  \end{equation}
  where $C_{\mu\nu}=\epsilon\indices{_\mu^{\alpha\beta}}\nabla_\alpha\left( R_{\beta\nu}-\frac14g_{\beta\nu}R\right)$ is the Cotton tensor who is symmetric, traceless and identically conserved. Moreover, it vanishes identically for GR solutions. Therefore all GR solutions are solutions of TMG. The conserved charges, the mass and angular momentum associated to the Killing vectors $\partial_t$ and $\partial_\phi$ respectively, are modified by the presence of the Chern-Simons term. Their expressions can be found using the general covariant phase space method of \cite{Barnich:2001jy, Barnich:2003xg, Barnich:2007bf} or, because we are dealing with exact Killing vectors, in the ADT formalism \cite{Abbott:1981ff,Iyer:1994ys, Bouchareb:2007yx,Compere:2008cv}. Their derivations are implemented in the "Surface Charge" package\footnote{G. Comp\`ere, \url{http://www.ulb.ac.be/sciences/ptm/pmif/gcompere/package.html}}. 
The entropy is also modified \cite{Solodukhin:2005ah, Kraus:2005zm, Bouchareb:2007yx, Tachikawa:2006sz}.
 Writing the metric in ADM form
\begin{equation} \label{ADM}
ds^2= -N(r)^2 dt^2+ \frac{dr^2}{f(r)^2}+ R(r)^2 (N^{\phi}(r) dt + d\phi)^2,
\end{equation}
the entropy in TMG is given by 
\begin{equation}
S^{TMG}_\pm=\frac\pi 2 R(r_\pm) -\left. \frac \pi {2\mu} \frac {R(r)^2 f(r) N^{\phi}(r)'}{2 N(r)} \right|_{r=r_\pm}.
\end{equation}

In order for BTZ to satisfy the equations of motion, the cosmological constant has to be $\Lambda=-1$ (we set $l=1$) while the Chern-Simons coupling can be arbitrary. The non extremal BTZ metric is given by 
\begin{equation}
  g_{\mu\nu}=
  \left(
  \begin{array}{ccc}
     \left(8 M-r^2\right) & 0 &4   J \\
   0 & \frac{1 }{\frac{16 J^2}{r^2}+r^2-8 M} & 0 \\
   4   J & 0 &   r^2
  \end{array}
  \right)
\end{equation}
with $M^2>J^2$. This condition guarantees the absence of naked singularities.
The ADM functions defined in \eqref{ADM} for BTZ are
\begin{equation}
N(r)^2=f(r)^2= \frac{16 J^2}{r^2}+r^2-8 M \quad , \quad N^{\phi}(r)=\frac{4J}{r^2} \quad , \quad R(r)^2=r ^2.
\end{equation}

\subsection{Thermodynamic quantities}
The horizons are located at
\begin{equation}
r_\pm=2\sqrt{M\pm\sqrt{M^2-J^2}}.
\end{equation}
We express the thermodynamic quantities 
in terms of the BTZ conserved charged in GR $(M,J)$ and also in terms of the horizons radii $(r_\pm)$.
\begin{equation} \label{BTZparam}
\begin{array}{c|c|c|c|c|c}
 & M^{TMG} &J^{TMG}& S^{TMG} & T & \Omega \\
 \hline
&&&&&\\
(M,J) & M+\frac J \mu& J+\frac M \mu  & \pi  \sqrt{\sqrt{M^2-J^2}+M} & \frac{2 \sqrt{M^2-J^2}}{\pi  \sqrt{\sqrt{M^2-J^2}+M}}  & \frac{J}{\sqrt{M^2-J^2}+M} \\
&&& \quad+\frac{\pi  J}{\mu  \sqrt{\sqrt{M^2-J^2}+M}} && \\
&&&&&\\
(r_+,r_-) & \frac{r_+^2+r_-^2}{8}+\frac{r_+ r_-}{4\mu}  & \frac{r_+ r_-}{4}+\frac{r_+^2+r_-^2}{8\mu} & \frac {\pi r_+}2+\frac {\pi r_-}{2\mu} & \frac{r_+^2-r_-^2}{2\pi r_+}  & \frac{r_-}{r_+} \\
\end{array}
\end{equation}
The temperature and the angular velocity are not modified by the presence of the Chern-Simons term since they depend only on the metric. 
These quantities satisfy the first law of black hole thermodynamics, namely
\begin{equation}
dM=T dS + \Omega dJ.
\end{equation}

%

\subsection{Stability and Hawking-Page transition}
We study the stability in the grand canonical ensemble, i.e. at fixed temperature and angular velocity. The relevant thermodynamic potential is the Gibbs free energy $G(T,\Omega)$. In gravity theories it is given by \refe{freeen}:
\be
   G(T,\Omega) = T S[g_c]
\ee
where $g_c$ is a euclidean saddle with boundary torus of modular parameter $\tau = \frac{1}{2 \pi} (-\beta \Omega_E + i \frac{\beta}{l})$. Evaluating $S[g_c]$ can be a tricky procedure, even more subtle in TMG that the action is not diffeomorphism-invariant. For asymptotically AdS$_3$ spaces however, the result is known \cite{Kraus:2006wn}. At fixed potentials $(T,\Omega)$, there are at least two geometries contributing to the partition function in a saddle point approximation: thermal AdS$_3$ and rotating BTZ with parameters $(r_+,r_-)$ given by \refe{BTZparam}. The result for thermal AdS$_3$ is
\begin{equation}\label{actioneuclads}
S_E[AdS(\tau)]=\frac{i \pi}{12\, l}\left( c\tau-\tilde c\bar \tau\right )
\end{equation}
where $c,\tilde c$ are the central charges of the theory and $\bar{\tau}$ is the complex conjugate of $\tau$.
In TMG, they are 
\begin{equation}
c,\tilde c=\frac{ 3\, l}{2}\left( 1\pm \frac1{\mu\, l} \right) .
\end{equation}
So, in TMG, we have
\begin{equation}
S_E[AdS(\tau)]=-\frac{1}{8\,l\,T}\left( 1- \frac{\Omega_E}{ \mu}\right ),
\end{equation}
and the Gibbs free energy is
\begin{equation}
G_{AdS}=-\frac1{8\,l}\left(1-\frac{\Omega_E}{\mu}\right). 
\end{equation}
For rotating BTZ, the usual trick is to notice that BTZ is equivalent to thermal rotating AdS in certain coordinates. The details are reviewed in Appendix 2. We have that
\begin{equation}
ds^2_{BTZ}\left [-\frac1\tau\right ]=ds^2_{AdS}[\tau]
\end{equation}
and hence the Euclidean on-shell action for BTZ is 
\begin{equation}\label{actioneuclbtz}
S_E[BTZ(\tau)]=-\frac{i \pi}{12 \, l}\left( \frac c{\tau}-\frac{\tilde c}{\bar \tau}\right ).
\end{equation}
In TMG, we get 
\begin{equation}
S_E[BTZ(\tau)]=-\frac{l^2\,\pi^2T}{2(1+l^2\,\Omega_E^2)}\left(1-\frac{\Omega_E }{\mu}  \right)  
\end{equation}
and hence the free energy.

It is important to notice that these derivations are very specific to AdS$_3$ spaces and could not systematically be used for spaces with different asymptotics. One can however proceed indirectly. Indeed, the first law of thermodynamics can be integrated to yield
\begin{equation}\label{freeenergy}
G=M-T S - \Omega J.
\end{equation}
For BTZ and AdS, using \refe{BTZparam} and the fact that the AdS$_3$ vacuum has $M=-1/8$, $J=0$ and zero entropy in the classical limit,
we get
\begin{align}\label{GBTZTMG}
& G^{BTZ}(T,\Omega)=-\frac{\pi ^2 T^2}{2(1-\Omega ^2)} \left(1+ \frac{\Omega}{\mu} \right)\\\label{GAdSTMG}
& G^{AdS}(T,\Omega)=-\frac18\left( 1-\frac{\Omega}{\mu}\right) 
\end{align}
which coincides with the above results obtained in euclidean signature.\footnote{This requires to analytically continue the Chern-Simons coupling.}

%



\subsubsection{Local Stability}
We are asking if the considered phase is locally stable, i.e. under small perturbations. 
The second law of thermodynamics states the stability criteria, namely a stable system satisfies $\Delta S\leq0$ where $\Delta S$ is the variation of the entropy. As we are considering small perturbations, we develop $\Delta S$ in Taylor series $\Delta S=\delta S+\delta^2S+\delta^3S+...$. Therefore, at thermodynamic equilibrium ($\delta S=0$), a system will be stable if and only if $\delta S^2\leq0$. It is easy to adapt it for the considered ensemble. For example, in the grand canonical ensemble, the stability condition becomes the requirement for a system to have the Hessian $H$ of its free energy $G(T,\Omega)$ negative definite, i.e. the eigenvalues of 
\begin{equation}
H=\begin{pmatrix}
 \frac{\partial^2 G}{\partial T^2} & \frac{\partial^2 G}{\partial T \partial \Omega} \\
\frac{\partial^2 G}{\partial \Omega \partial T} &\frac{\partial^2 G}{\partial \Omega^2}
\end{pmatrix}
\end{equation}
are negative. 

For BTZ at fixed temperature and angular velocity in TMG, we get the following conditions
\begin{equation}
\Omega^2<1 \quad , \quad \mu^2>1 . 
\end{equation}
The first condition is just the non extremality condition. Therefore, BTZ black holes are locally stables for TMG with $\mu>1$. Note that GR case is consistently included: the non extremal BTZ black holes are indeed locally stable in GR. 
Moreover, $\mu=1$ and $\Omega=1$ define the boundaries of the stability area, they are called the spinodal curve.

\subsubsection{Global Stability and Modular Invariance}

The question of global stability is to determine which phase among the possible ones is the most likely.
Here, for each $\mu$ bigger than $1$\footnote{In order for the system to still be locally stable. It is meaningless to talk about a phase who is globally stable but not locally. }, we have two classical solutions contributing as classical saddle points: the BTZ black hole and rotating thermal AdS both at modular parameter $\tau$.  \footnote{We discard the other Euclidean solutions of \cite{Maldacena:1998bw, Maloney:2007ud} for the moment since they don't have a well defined Lorentzian continuation.} 
In the classical limit, the partition function reduces then to
\begin{equation}
Z(\tau)=e^{-S_E[AdS(\tau)]}+e^{-S_E[BTZ(\tau)]},
\end{equation}
so we need to compare the free energies of the two possible phases.

The difference in free energies is
\begin{equation}
\Delta G= G_{AdS}-G_{BTZ}
=\left( -\frac18+\frac{\pi^2 T^2}{2(1-\Omega^2)}\right) +\frac\Omega\mu\left( \frac{1}{8}+\frac{\pi^2T^2}{2(1-\Omega^2)}\right). 
\end{equation}
 When $\Delta G>0$ ($<0$), it means that BTZ (AdS) is the dominant phase. The second term due to Chern-Simons term cancels in the absence of rotation. Therefore the phase diagram will be same as in pure GR, the Hawking-Page transition occurring at the self-dual temperature \refe{sdbeta}.

\begin{figure}[h]
        \centering
        \begin{subfigure}[t]{0.32\textwidth}
                \includegraphics[width=\textwidth]{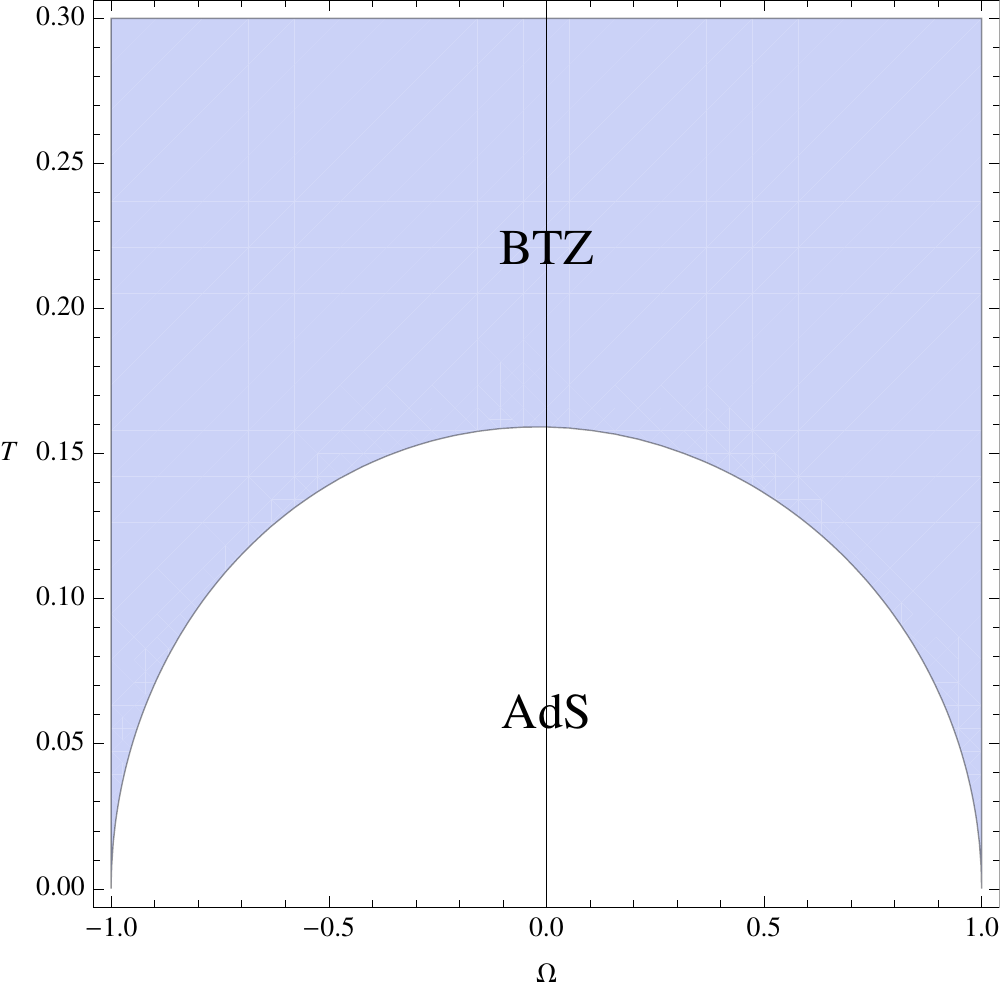}
                \caption{GR limit, e.g. $\mu=50$}
                \label{fig:gull}
        \end{subfigure}%
        ~ 
        \begin{subfigure}[t]{0.32\textwidth}
                \includegraphics[width=\textwidth]{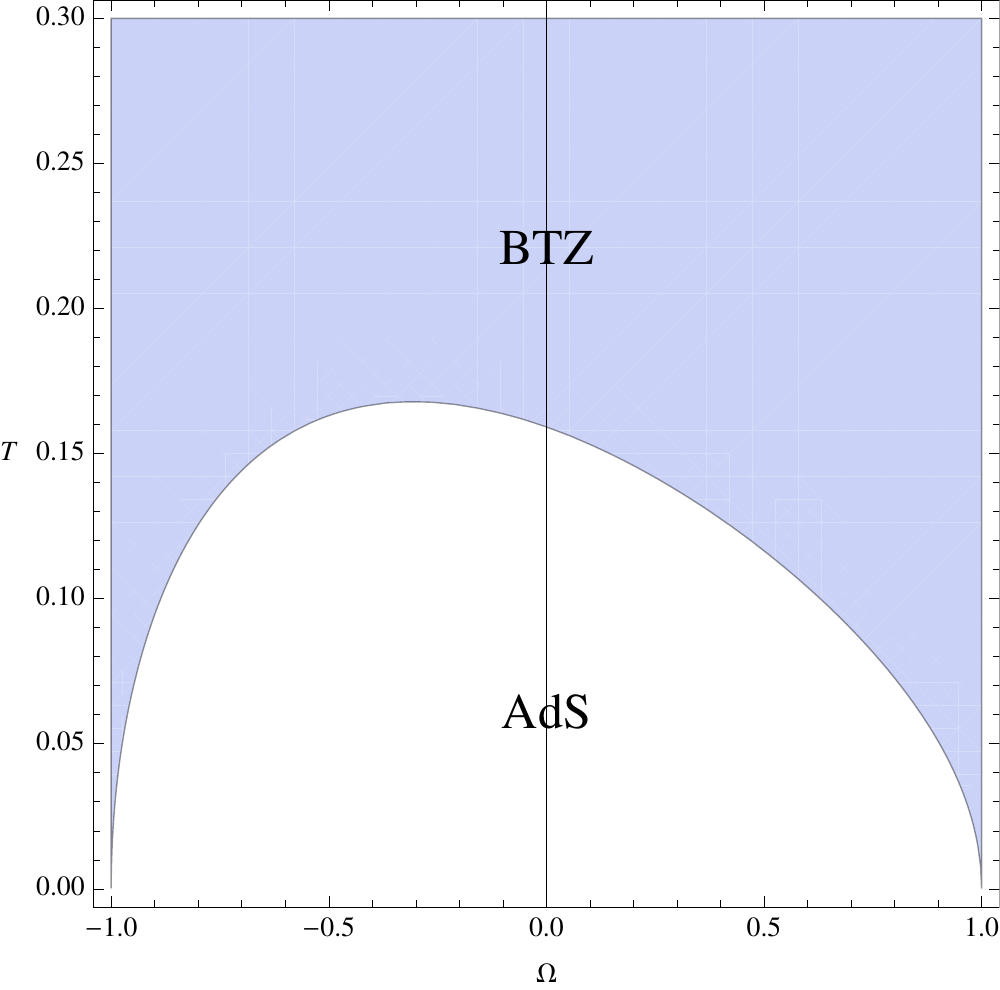}
                \caption{Stretched-squashed limit $\mu=3$}
                \label{fig:tiger}
        \end{subfigure}
        ~ 
        \begin{subfigure}[t]{0.32\textwidth}
                \includegraphics[width=\textwidth]{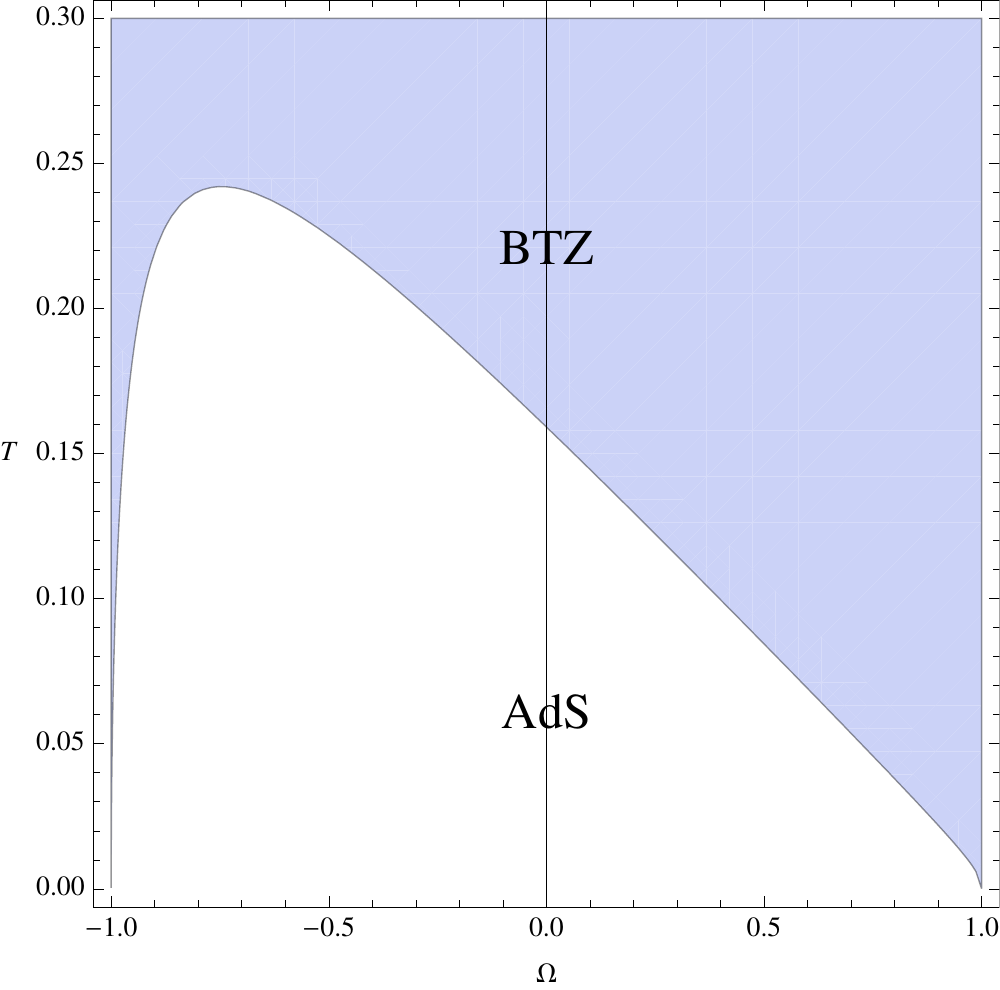}
                \caption{Local stability limit $\mu\lesssim1$}
                \label{fig:mouse}
        \end{subfigure}
        \caption{BTZ-AdS phase diagram for different Chern-Simons couplings.   BTZ dominates in the purple section and AdS in the white. }\label{fig:BTZ-AdS-TMG}
\end{figure}

With rotation, the Chern-Simons contribution does not factorise and has a impact of the phase transition behaviour. 
As $\mu$ decreases from $\infty$ (pure GR), the Chern-Simons contribution becomes bigger and we observe a growing asymmetry between positive and negative angular velocities. It is the signature of the parity breaking of the Chern-Simons term. 
Graphically, the situation is presented in figure \ref{fig:BTZ-AdS-TMG} for different values of $\mu$. 

\section{Phase Transitions in WAdS$_3$}\label{wbtztmg}

We now turn to the analysis of the Warped black holes introduced in \refe{WBTZ}. Their metric reads
  \begin{equation}\label{WBTZ-metric}
     g_{\mu\nu}=
  \left(
  \begin{array}{ccc}
   -r^2-\frac{H^2 \left(-r^2-4 J+8 M\right)^2}{4 (M-J)}+8 M & 0 & 4 J-\frac{H^2 \left(4 J-r^2\right)
     \left(-r^2-4 J+8 M\right)}{4 (M-J)} \\
   0 & \frac{1}{\frac{16 J^2}{r^2}+r^2-8 M} & 0 \\
   4 J-\frac{H^2 \left(4 J-r^2\right) \left(-r^2-4 J+8 M\right)}{4 (M-J)} & 0 & r^2-\frac{H^2 \left(4
     J-r^2\right)^2}{4 (M-J)}
  \end{array}
  \right).
   \end{equation}
   
   In order to satisfy the equations of motion of TMG, the cosmological constant $\Lambda$ and the Chern-Simons couping $\mu$ have to be 
   \begin{equation} \label{muandH}
   \mu=3\sqrt{1-2 H^2}  \quad , \quad \Lambda=-\frac{3+2 H^2}{3}. 
   \end{equation}
The range of allowed values for $H^2$ to keep $\mu$ real and $\Lambda$ negative is
\begin{equation}
-\frac32<H^2<\frac12.
\end{equation}

   In ADM formalism, the functions appearing in \eqref{ADM} are
  \begin{align}
  & \nonumber
   N(r)^2=\frac{4 \left(2 H^2-1\right) \left(-16 J^3+16 J^2 M+J \left(8 M r^2-r^4\right)+M r^2 \left(r^2-8
      M\right)\right)}{H^2 \left(r^2-4 J\right)^2+4 r^2 (J-M)}\\
  & \nonumber
  f(r)^2=\frac{16 J^2}{r^2}-8 M+r^2 \\ \nonumber
  & N^{\phi}=\frac{-16 \left(H^2-1\right) J^2+16 \left(2 H^2-1\right) J M+H^2 r^2 \left(r^2-8 M\right)}{H^2
     \left(r^2-4 J\right)^2+4 r^2 (J-M)} \\
  & R(r)^2=\frac{4 r^2 (M-J)-H^2 \left(r^2-4 J\right)^2}{4 (M-J)}.
  \end{align}
  The horizons are located in
   \begin{equation}
   r_\pm= 2 \sqrt{M\pm \sqrt{-J^2+M^2}}
   \end{equation}
 and the thermodynamic quantities are
\begin{equation}
\begin{array}{c|c|c|c|c}
  M^{TMG} &J^{TMG}& S^{TMG} & T & \Omega \\
 \hline
 \frac{\left(3-4 H^2\right) M+J}{3 \sqrt{1-2 H^2}}&
\frac{\left(3-4 H^2\right) J+M}{3 \sqrt{1-2 H^2}} &
\frac{\pi  \left(J-\left(4 H^2-3\right) \left(\sqrt{M^2-J^2}+M\right)\right)}{3 \sqrt{\left(1-2
    H^2\right) \left(\sqrt{M^2-J^2}+M\right)}}&
  \frac{2 \sqrt{M^2-J^2}}{\pi  \sqrt{\sqrt{M^2-J^2}+M}} & 
   \frac{J}{\sqrt{M^2-J^2}+M} \\
\end{array}
\end{equation}  
Notice that the thermodynamic potentiels are not affected by the deformation.

As argued in Sect.~\ref{sectthermal}, the global ground state corresponds to the above geometry with $M=-\frac18$ and $J=0$, referred to as Timelike WAdS$_3$ because the norm of the (real) deformation vector now has become negative.
%
 Its metric is
\begin{equation}
 g = 
\left(
\begin{array}{ccc}
 -r^2+2 H^2 \left(-r^2-1\right)^2-1 & 0 & -2 H^2 r^2
   \left(-r^2-1\right) \\
 0 & \frac{1}{r^2+1} & 0 \\
 -2 H^2 r^2 \left(-r^2-1\right) & 0 & r^2\left(1+2 H^2 r^2\right)
\end{array}
\right).
\end{equation}
%

We now reproduce the discussion we presented for BTZ, avoiding the caveat of having to compute on-shell actions in TMG by evaluating \refe{freeenergy} directly, and discuss the issue of stability and Hawking-Page transition. To this end we compute the free energies and obtain
\begin{align}\label{GWBTZL}
& G_{WAdS}(T,\Omega)=-\frac{3-4 H^2-\Omega}{24 \sqrt{1-2 H^2}}\\\label{GWAdSL} 
& G_{WBTZ}(T,\Omega)=-\frac{\pi ^2 T^2 \left(3-4 H^2+\Omega \right)}{6 \sqrt{1-2 H^2} \left(1-\Omega ^2\right)}.
 \end{align}

\subsection{Local Stability}
The local stability implies that
\begin{equation}
\Omega^2<1  \quad , \quad H^2<\frac12 .
\end{equation} 
As $H^2$ and $\mu$ are related by \eqref{muandH}, the second condition implies $\mu^2>0$. 
Note that the BTZ case ($H^2=0$) in TMG is consistently included. Therefore, in the ensemble considered, the Warped black holes are always locally stable.

\subsection{Global Stability and Modular Invariance}  
There is a priori no obvious counterpart of expression \refe{actioneuclads} for the thermal WAdS$_3$ geometry. However, $\beta G_{WAdS}(T,\Omega)$ evaluated here above can be rewritten in a suggestive way (after continuing $\Omega$ to euclidean signature) as
\be \label{GWADS}
  \beta G_{WAdS}(T,\Omega) = \frac{i \pi}{12\, l}\left( c_R  \tau- c_L \bar \tau \right )
\ee
where $c_{L/R}$ were defined through \refe{warpedclcr} and are given by
\be
   c_R=\frac{2\,l}{\sqrt{1-2H^2}}\left (1-H^2\right ), \quad c_L=l \sqrt{1-2H^2}.
\ee
Expression \refe{GWADS} is the would-be Euclidean on-shell action for the WAdS thermal background.
Now, using the same change of coordinate as the one mapping thermal AdS$_3$ on BTZ we observe (see Appendix 2) that \begin{equation} \label{wbtzwads}
ds^2_{WBTZ}\left [-\frac1\tau\right ]=ds^2_{WAdS}[\tau].
\end{equation}
Therefore, one expects  
\be \label{GWBTZ}
  \beta G_{WBTZ}(T,\Omega) = -\frac{i \pi}{12 \, l}\left( \frac{c_R}{\tau}-\frac{c_L}{\bar \tau} \right ).
\ee
This is found to coincide with the result from the integrated first law \refe{GWAdSL}. Similarly to the situation in AdS$_3$ \cite{Maldacena:1998bw}, this suggests the existence of an SL(2,$\mathbf{Z}$) family of Warped black holes.


%

To establish the phase diagram, we now compare the two free energies \eqref{GWBTZL}\eqref{GWAdSL}.
In the case without rotation, we have the same behaviour as BTZ in TMG because the deformation parameter appears in the same global factor in the two free energies.  
So, the phase transition behaviour is independent of the deformation parameter.
 
\begin{figure}[h]
         \centering
         \begin{subfigure}[h]{0.32\textwidth}
                 \includegraphics[width=\textwidth]{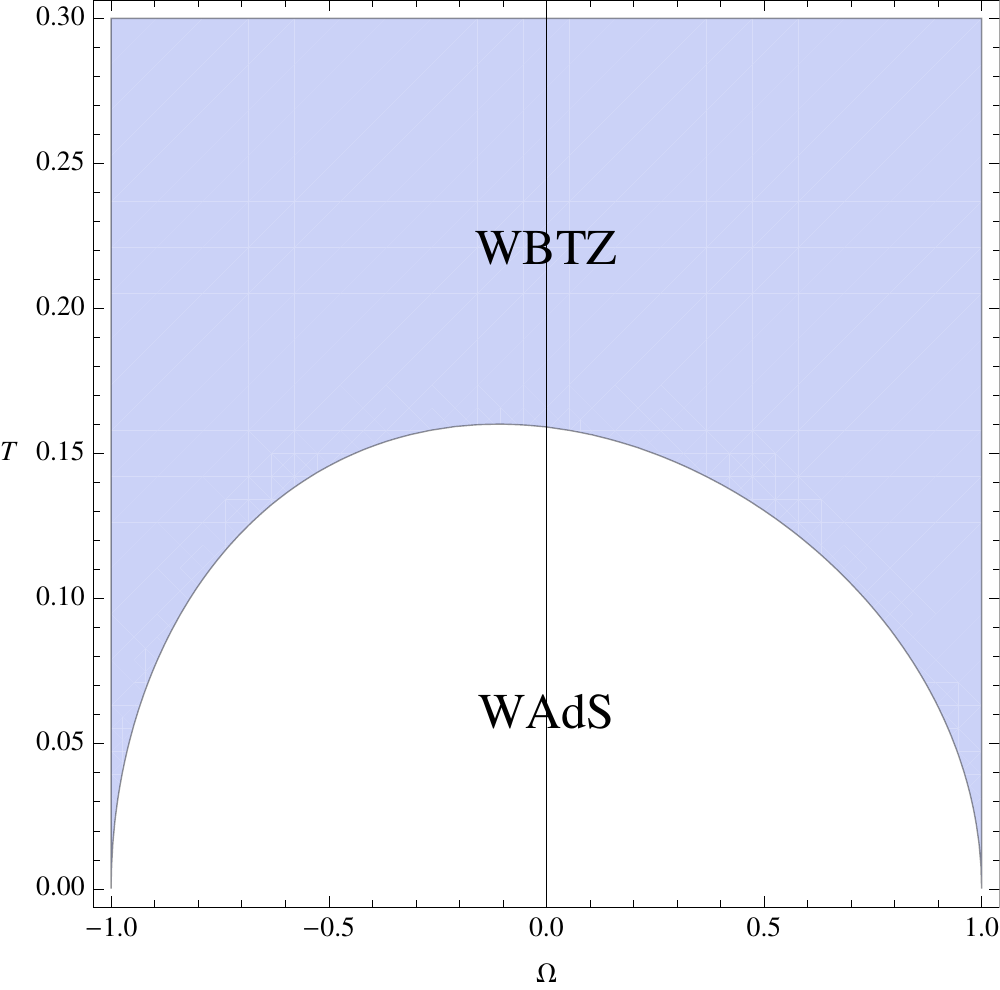}
                 \caption{Flat space limit $H^2=-\frac32$ or $\mu=36$ and $\Lambda\lesssim0$}
         \end{subfigure}%
         ~ 
         \begin{subfigure}[h]{0.32\textwidth}
                 \includegraphics[width=\textwidth]{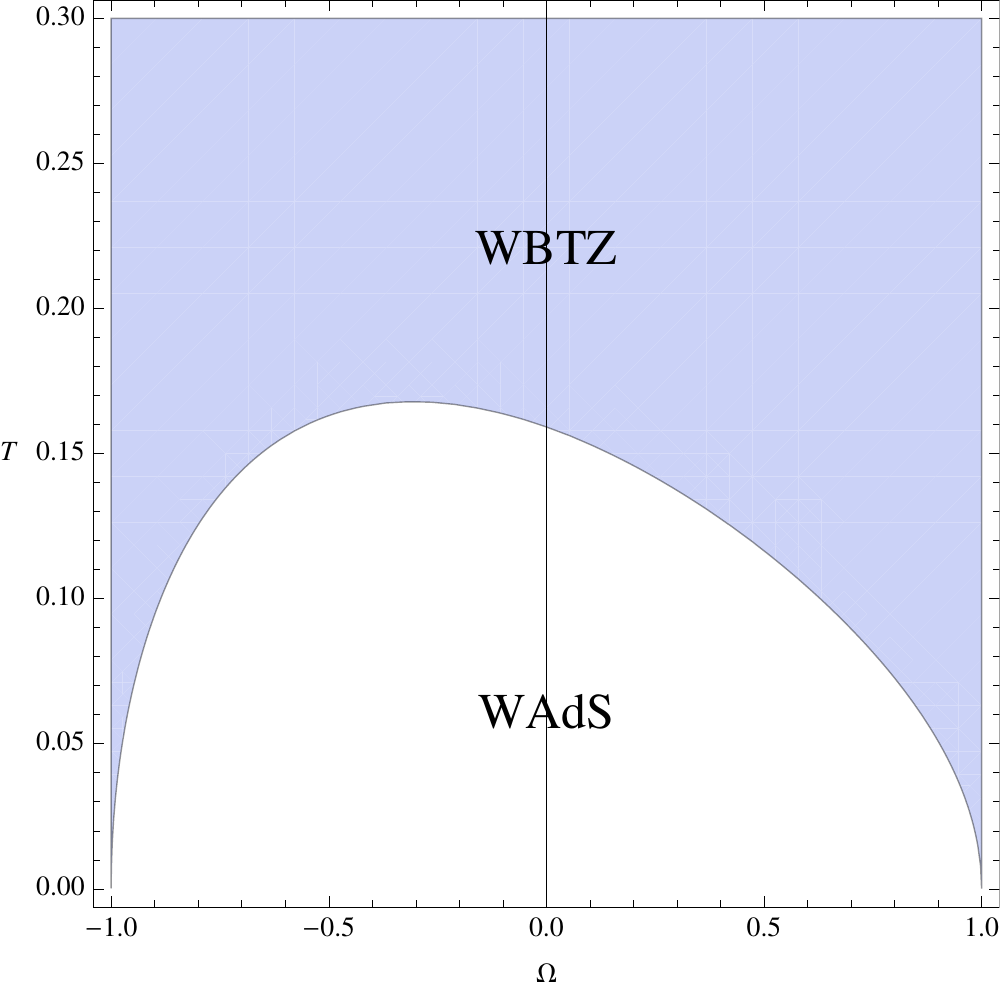}
                 \caption{Stretched-squashed limit $H^2=0$ or $\mu=3$ and $\Lambda=-1$}
         \end{subfigure}
         ~ 
         \begin{subfigure}[h]{0.32\textwidth}
                 \includegraphics[width=\textwidth]{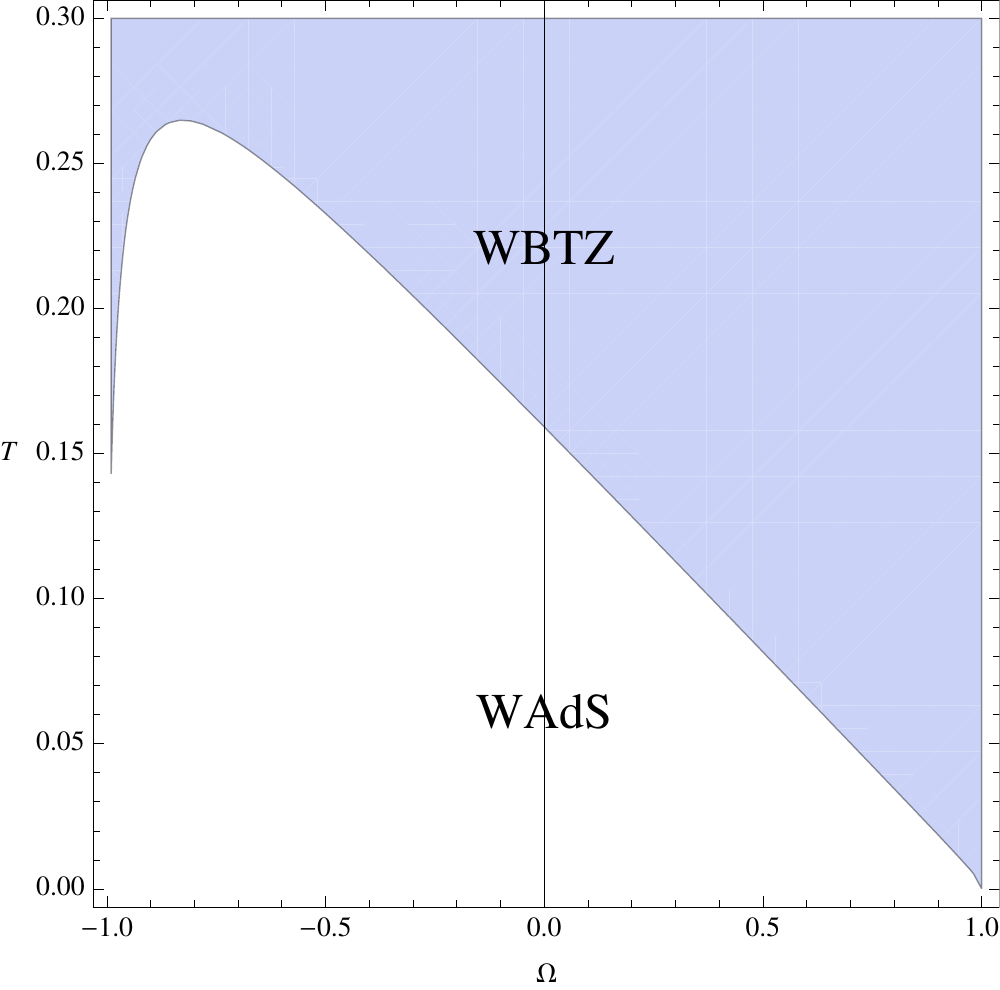}
                 \caption{Local stability limit $H^2\lesssim \frac12$ or $\mu\gtrsim0$ and $\Lambda\lesssim-2/3$}
         \end{subfigure}
         \caption{WBTZ-WAdS phase diagram for different deformation parameter values.   WBTZ dominates in the purple section and WAdS in the white. }\label{fig:WBTZ-WAdS}
 \end{figure}

In the general case, there is an asymmetry in the angular velocity due to the Chern-Simons term.
We have the same qualitative behaviour than the BTZ-AdS case in TMG. 
As $H^2$ becomes bigger, or equivalently $\mu$  becomes smaller, the asymmetry is larger. 
Graphically, the situation is present in figure \ref{fig:WBTZ-WAdS} for different values of $\mu$. 

\section{A digression on Inner Horizons}\label{SectInner}

A priori, we could think that what happens inside a black hole has no repercussion whatsoever on the dynamics outside. Nevertheless, it was argued that inner horizon thermodynamic quantities combined with outer horizon ones can for instance give us information about the scattering data around the black hole \cite{Castro:2013kea,Chen:2012mh,Chen:2013rb}. In the previous section, we have shown that the warped black hole and background free energies could be expressed in a simple way in terms of the thermal torus parameter $\tau$ and the parameters $c_L$ and $c_R$ defined in \refe{warpedclcr}. While the latter can be obtained in general from asymptotic symmetries \cite{Compere:2008cv}, $c_L$ instead depends on the vacuum background charges and is not fixed by symmetries alone. In this section, we show how both values can be derived from inner and outer horizon properties, following \cite{Chen:2013aza, Chen:2013qza, Chen:2012mh}.


\subsection{Inner horizon thermodynamic quantities}
The thermodynamic quantities for the BTZ and the WBTZ black holes at the inner horizon are computed below:
\begin{equation}
\begin{array}{c|c|c|c}
& S^{TMG}_- & T_- & \Omega_- \\
 \hline
&&&\\
BTZ(M,J) & \pi  \sqrt{M-\sqrt{M^2-J^2}} +\frac{\pi  J}{\mu  \sqrt{M-\sqrt{M^2-J^2}}} & \frac{2 \sqrt{M^2-J^2}}{\pi  \sqrt{M-\sqrt{M^2-J^2}}}  & \frac{J}{M-\sqrt{M^2-J^2}} \\
&&&\\
BTZ(r_+,r_-) & \frac {\pi r_-}2+\frac {\pi r_+}{2\mu} & \frac{r_+^2-r_-^2}{2\pi r_-}  & \frac{r_+}{r_-} \\
&&&\\
WBTZ(M,J) &\frac{\pi  \left(\left(4 H^2-3\right) \left(\sqrt{M^2-J^2}-M\right)+J\right)}{3 \sqrt{\left(2 H^2-1\right) \left(\sqrt{M^2-J^2}-M\right)}} &\frac{2 \sqrt{M^2-J^2}}{\pi  \sqrt{M-\sqrt{M^2-J^2}}}    &\frac{J}{M-\sqrt{M^2-J^2}} \\
\end{array}
\end{equation}
We denote the quantities related to the inner and outer horizon with a index $-$ and $+$ respectively. 
We note the following symmetry between the outer and inner horizon quantities:
\begin{equation}
S_+=\left. S_-\right|_{r_+ \leftrightarrow r_-} \quad,\quad T_+=- \left. T_-\right|_{r_+ \leftrightarrow r_-}  \quad,\quad \Omega_+=\left. \Omega_-\right|_{r_+ \leftrightarrow r_-}.
\end{equation}
This symmetry implies that the inner quantities also satisfy at a first law of thermodynamics\footnote{In general, it was showed that under reasonable assumptions, the first law of thermodynamics at the outer horizon implies the one at the inner horizon \cite{Chen:2012mh}.} as noted previously in \cite{Larsen:1997ge, Cvetic:2010mn, Castro:2012av, Ansorg:2008bv, Curir:1981uc}
\begin{equation}
dM=-T_-dS_-+\Omega_-dJ.
\end{equation}

\subsection{Central and vacuum charges from inner thermodynamics}
We now derive an expression for the quantities $c_L$ and $c_R$ from the horizons thermodynamics. As we previoulsy emphasized, only $c_R$ is interpreted as a Virasoro central charge, while $c_L$ is just a notation for the quantity \refe{warpedclcr}, which we will still call ``central charge" by convention.

For theories without gravitational anomalies, it was highly suggested that the temperatures and the central charges of the dual field theories can be rederived in scattering data. The properties of the Klein-Gordon equation in a black hole background, exhibiting a hidden conformal symmetry \cite{Castro:2010fd}, are highlighted by the techniques of monodromies \cite{Castro:2013kea}. They turn out to be closely related to inner and outer horizon thermodynamics. We extend the results of \cite{Castro:2013kea} in the case with gravitational anomalies as TMG. 

Left and right temperatures and entropies are defined as
 \begin{equation}
T_{R,L}=\frac{T_-\pm T_+}{\Omega_--\Omega_+}
\end{equation}
and
\begin{equation}
S_\pm=S_R\pm S_L.
\end{equation}
We have $S_{R,L}\propto T_{R,L}$ for BTZ and WBTZ black holes. To prove this fact, we consider the product of the horizon entropies. It can be expressed in terms of the conserved charges $Q$ associated with $\partial_t$ and $\partial_\phi$,
\begin{equation}\label{SiSo}
S_+S_-=G( Q_{\partial_t},Q_{\partial_\phi}).
\end{equation}
We take the variation of this equation. We express the left hand side in terms of $S_L,S_R$ and their variations $\delta S_L,\delta S_R$,
\begin{equation}
\delta(S_+ S_-)=2(S_R\delta S_R-S_L \delta S_L).
\end{equation}
The variation of the right hand side is 
$\frac{\delta G}{\delta Q_{\partial_t}}\delta Q_{\partial_t}+\frac{\delta G}{\delta Q_{\partial_\phi}}\delta Q_{\partial_\phi}$. 
Now, the inner and outer first laws read\cite{Wald:1993nt, Castro:2013kea}
\begin{equation}
\delta S_\pm=\pm \frac{1}{T_\pm}(\delta Q_{\partial_t} -\Omega_\pm\delta Q_{\partial_\phi}).
\end{equation}
We inverse this equation and obtain the variation of the charges in terms of $\delta S_\pm$ and finally in terms of $\delta S_{R,L}$. We get
\begin{align}
& \delta Q_{\partial_t} = \delta S_R\left( \frac{T_R(\Omega_-+\Omega_+)+T_L(\Omega_+-\Omega_-)}2 \right)
 -\delta S_L\left( \frac{T_R(\Omega_+-\Omega_-)+T_L(\Omega_++\Omega_-)}2 \right)\\
& \delta Q_{\partial_\phi} = \delta S_R T_R - \delta S_L T_L.
\end{align}
The variation of \eqref{SiSo} is
\begin{align}
S_R\delta S_R-S_L \delta S_L= &
\left( \frac14 \frac{\delta G}{\delta Q_{\partial_t}} \left( \frac{T_L}{T_R}(\Omega_+-\Omega_-)+(\Omega_-+\Omega_+)\right) T_R
+\frac1{2} \frac{\delta G}{\delta Q_{\partial_\phi}} T_R \right) \delta S_R\\
&-\left( \frac14 \frac{\delta G}{\delta Q_{\partial_t}} \left( \frac{T_R}{T_L}(\Omega_+-\Omega_-)+(\Omega_-+\Omega_+)\right) T_L
+\frac1{2} \frac{\delta G}{\delta Q_{\partial_\phi}} T_L \right)\delta S_L.
\end{align}
So we get 
\begin{align}
& S_R=\left( \frac14 \frac{\delta G}{\delta Q_{\partial_t}} \left( \frac{T_L}{T_R}(\Omega_+-\Omega_-)+(\Omega_-+\Omega_+)\right) 
+\frac12 \frac{\delta G}{\delta Q_{\partial_\phi}} \right) T_R  \\
& S_L=\left( \frac14 \frac{\delta G}{\delta Q_{\partial_t}} \left( \frac{T_R}{T_L}(\Omega_+-\Omega_-)+(\Omega_-+\Omega_+)\right) 
+\frac12 \frac{\delta G}{\delta Q_{\partial_\phi}}  \right) T_L.
\end{align}
It turns out that the quantities $S_{R,L}/T_{R,L}$ usually happen to be constant, which are conventionally denoted by $\frac{\pi^2 l}{3} c_{R,L}$.
We then obtain 
\be\label{Srl}
S_{R,L}=\frac{\pi^2 l}{3} c_{R,L} T_{R,L}
\ee
justifying the name ``central charges" for $c_{R,L}$ from the similarity between \refe{Srl} and 2d CFT entropies.

It is more convenient to express the central charges in terms of the horizons quantities, they are
\begin{equation}
 c_{R,L}=\frac{3}{2\pi^2}\left( \frac{\delta( S_+S_-)}{\delta Q_{\partial_\phi}}  + \frac{\delta (S_+S_-)}{\delta Q_{\partial_t} } \left( \frac{\Omega_+ T_-\pm\Omega_- T_+}{(T_-\pm T_+)} \right) 
\right)  .
\end{equation}
For BTZ, we have right and left temperatures given by
\begin{equation}
T_{R,L}=\frac{r_+\pm r_-}{2 \pi }.
\end{equation} 
The product of horizon entropies is
\begin{equation}
S_+S_-= \pi^2 \left ( J^{TMG}+\frac{M^{TMG} }{\mu }\right ).
\end{equation}
The central charges are
\begin{equation}
c_{R,L}=\frac{3 l}{2 G}\left( 1\pm \frac1{\mu l}\right).
\end{equation}
For WBTZ, the right and left moving temperature are the same as BTZ. 
The product of horizon of the entropies is
\begin{equation}
S_+S_-= -\frac{\pi ^2 \left(\left(4 H^2-3\right) J^{WBTZ}-M^{WBTZ}\right)}{3 \sqrt{1-2 H^2}}.
\end{equation}
The central charges are
\begin{align}
& c_R=\frac{2\,l}{G} \frac{1-H^2}{\sqrt{1-2H^2}} \\
& c_L=\frac{l}{G} \sqrt{1-2H^2}.
\end{align}
Performing the change of variables (see Appendix 1) to the usual system of coordinates for the warped black holes. We get the following central charges
\begin{equation}
c_R = \frac{\hat l\,(5\nu^2 + 3)}{G\nu(\nu^2+3)} \quad , \quad c_L = \frac{4\nu\,\hat l}{G(3+\nu^2)}
\end{equation}
which are exactly the ones found in \cite{Anninos:2008fx}.


\section{Outlook}

The aim of this paper was to study phase transitions for asymptotically WAdS$_3$ black holes. The latter have been conjectured to be dual to certain thermal density matrices in a 2d WCFT. After reviewing the Hawking-Page phase transition for BTZ black holes in TMG and its relation to modular invariance of the dual holographic CFT$_2$, we turned to WAdS$_3$ black holes viewed as UV deformations of BTZ. Comparing the Gibbs free energies of the black hole and the WAdS thermal background in a given ensemble, we observed a phase transition analogous to the one present in AdS$_3$: for high temperatures, the partition function is dominated by the black hole phase, while at low temperatures the dominant one is a thermal gaz in WAdS$_3$.  The transition was found to occur along a curve in the $(\beta, \Omega)$ plane defined by
\be
  c_R (\tau + \frac{1}{\tau}) - c_L (\bar \tau + \frac{1}{\bar \tau}) = 0.
\ee
Considering the solution $\tau = - \frac{1}{\tau}$ and $\bar \tau = - \frac{1}{\bar \tau}$ corresponds to putting the theory at vanishing angular potential and yields the self-dual temperature \refe{wsdtemp} of the WCFT partition function:
\be
  \beta_{HP} = \beta_{wsd} = 2 \pi.
\ee

We note that despite the deformation, the thermodynamic potentials $\beta$ and $\Omega$ remained unaltered compared to their BTZ expressions, and that the parameter $\tau$ parametrizing the boundary torus characteristic of a thermal theory on a circle was unchanged. The WCFT partition function being invariant under \refe{wmodularquad} and trivially under $\tau \ra \tau + 1$ due to the circle identifications is also therefore invariant under the whole SL(2,$\mathbf{Z})$ modular group, as already noted in \cite{Detournay:2012pc}. Moreover taking into account the SL(2,$\mathbf{Z})$ family of warped black holes suggested by \refe{wbtzwads} would yield in similar phase diagram as in AdS$_3$ gravity \cite{Maldacena:1998bw, Maloney:2007ud}

%

Let us conclude with a few additional comments. First, we want to comment on the relation between the present analysis and the work \cite{Birmingham:2010mj}. There the authors concluded that the warped black holes are locally unstable for all temperatures, in apparent contradiction with our results. This discrepancy is resolved by observing that the ensemble 
they considered, $Z(\beta_{bm}, \Theta_{bm}) = \mbox{Tr} e^{-\beta_{bm} P_0 + i \Theta_{bm} L_0}$, differs from the one we considered \refe{quadens} which was argued to be the one describing the warped black holes, see Sect.~\ref{sectthermal}. This leads to different conclusions regarding stability (actually, BTZ black holes would be unstable in the former ensemble). Second, we focused here on TMG at $\nu^2 \ne 1$. At the specific value $\nu^2 = 1$, spacelike warped black holes no longer exist and reduce to BTZ. However, another class of black holes and boundary conditions exist, containing the Null Warped spaces and black holes \cite{Anninos:2010pm}. The asymptotic symmetry group consists there of a chiral copy of a Virasoro algebra. Whether similar phase transitions are present is left as an open question.
Third, in view of \refe{wcftentropy} and \refe{wsdtemp}, which already inform us about  some of the properties a holographic WCFT should exhibit in the spirit of  \cite{Keller:2011xi, Hartman:2014oaa}, one cannot help but notice the striking ressemblance with the behaviour of a traditional holographic CFT$_2$. 
Actually, WAdS$_3$ spaces were originally proposed to be dual to a 2d CFT \cite{Anninos:2008fx}. However, the full conformal symmetry of WAdS$_3$ spaces in theories with purely gravitational degrees of freedom, if present, has not been uncovered so far. The situation is different when these spaces are embedded in string theory or when matter fields are present \cite{Guica:2011ia, Detournay:2012dz, Guica:2013jza}. 
In that case, it was shown that all consistent boundary conditions for AdS$_3$ could be mapped to boundary conditions for WAdS$_3$ through the introduction of an auxiliary, locally AdS metric constructed out of the original metric and matter fields. In particular, a phase space with 2 copies of a Virasoro algebra could be constructed \cite{Compere:2014bia}. On the other hand, recently Hofman and Rollier argued that the minimal setting to describe WCFTs holographically is a $SL(2,R) \times U(1)$ Chern-Simons theory \cite{Hofman:2014loa}, rather than a higher-curvature extension of Einstein-Hilbert theory like TMG. It would be interesting to reconsider our analysis in these two contexts. This could clarify the exact nature of field theories dual to these spaces, and the possible relation between CFTs and WCFTs, which ultimately could be relevant to Kerr/CFT.

\section*{Acknowledgements}

It is a pleasure to thank D. Birminghan and T. Hartman for useful correspondence, and G. Giribet for enlightening discussions. S.D and C.Z are supported in part by the ARC grant ``Holography, Gauge Theories and Quantum Gravity -- Building models of quantum black holes". S.D. is a Research Associate of the Fonds de la Recherche Scientifique F.R.S.-FNRS (Belgium).

\section*{Appendix 1: Coordinate systems for WAdS$_3$ Black Holes}\label{sec:coord}
 \addcontentsline{toc}{section}{Appendix 1: Coordinate systems for WAdS$_3$ Black Holes}
In this section, we present the change of coordinates and variables between two different metrics describing the spacelike stretched warped black holes: the metric \eqref{0807.3040BH} from \cite{Anninos:2008fx} and \eqref{WBTZ-metric}  from \cite{Detournay:2012pc}. 
The first metric is 
\begin{align} \nonumber
\frac{ds^2}{\h l^2}=
dT^2 & +\frac{d\hat {r}^2}{(\nu^2+3)(\hat r-\hat r_+)(\hat r-\hat r_-)}-\left( 2\nu\hat r -\sqrt{\hat r_+ \hat r_-(\nu^2+3)}\right) dT d\theta \\
&+\frac {\hat r} 4 \left( 3(\nu^2-1)\hat r+(\nu^2+3)(\hat r_++\h r_-)-4\nu\sqrt{\hat r_+ \hat r_-(\nu^2+3)  } \right)
\end{align}
with $\nu^2>1$ and $r_+>r_->0$ in order to describe non extremal black holes. The coordinates are $(T,\h r,\theta)$.
To be a solution of TMG equations of motion, we need 
\begin{equation}
\mu= \frac{3\nu} {\h l} \quad ,\quad \Lambda=-\frac1{\h l^2}. 
\end{equation} 
The second is the deformed BTZ metric given by \eqref{WBTZ-metric}
  \begin{equation}
     g_{\mu\nu}=
  \left(
  \begin{array}{ccc}
   -\frac{r^2}{l^2}-\frac{H^2 \left(-r^2-4l J+8l^2 M\right)^2}{4l^3 (lM-J)}+8 M & 0 & 4 J-\frac{H^2 \left(4l J-r^2\right)
     \left(-r^2-4 l J+8 l^2 M\right)}{4 l^2 (l M-J)} \\
   0 & \frac{1}{\frac{16 J^2}{r^2}+\frac{r^2}{l^2}-8 M} & 0 \\
   4 J-\frac{H^2 \left(4l J-r^2\right)
       \left(-r^2-4 l J+8 l^2 M\right)}{4 l^2 (l M-J)} & 0 & r^2-\frac{H^2 \left(4
     Jl-r^2\right)^2}{4l (lM-J)}
  \end{array}
  \right)
   \end{equation}
with $H^2<0$ and $ l /,M> |J|$. The coordinates are $(t,r,\phi)$.
To be a solution of the TMG equations of motion, we need 
\begin{equation}
\mu= \frac{3 \sqrt{1-2H^2}}{l} \quad ,\quad \Lambda=-\frac{3+2H^2}{3l^2}. 
\end{equation}

The change of coordinates and variables is given by
\begin{align}
& l=2 \frac{\h l}{\sqrt{3+\nu^2}}\\
& H^2=\frac 32 \frac{(1-\nu^2)}{(3+\nu^2)}\\
& M=\frac{\left(\nu ^2+3\right)^2 \left(\nu ^2 \left(2 \h r_-^2+\h r_- \h r_++2 \h r_+^2\right)-2 \nu  (\h r_-+\h r_+)
   \sqrt{\left(\nu ^2+3\right)\h r_-\h r_+}+3 \h r_- \h r_+\right)}{256 \h l ^2 \nu ^2}\\ 
& J= \frac{\left(\nu ^2+3\right)^{3/2} \left(\nu ^2 (\h r_+-\h r_-)^2-\left(\nu  (\h r_-+\h r_+)-\sqrt{\left(\nu ^2+3\right)
 \h r_- \h r_+}\right)^2\right)}{128 \h l  \nu ^2} \\
 & t = -\frac{2 \h l^2 T }{\sqrt{\nu ^2+3} \left(\nu  (\h r_-+\h r_+)-\sqrt{\left(\nu ^2+3\right)\h  r_-\h r_+}\right)}\\
& r = \frac{1}{4 \nu} \sqrt{\left(\nu ^2+3\right) \left(\nu ^2 (-3\h  r_- \h r_++4 \h r (\h r_-+\h r_+))-4 \nu  \h r \sqrt{\left(\nu
   ^2+3\right) \h r_- \h r_+}+3\h  r_-\h r_+\right)}\\
& \phi= \h l\left( -\theta + \frac T{\nu(\h r_-+\h r_+)-\sqrt{(3+\nu)\h r_-\h  r_+}}\right).
\end{align}

Introducing the $\hat u(1)$ level \cite{Compere:2008cv}
\be
 k = -\frac{\nu^2 + 3}{6 \nu}
\ee
and mass
\be
  Q_{\p_T} =: P_0 = \frac{\nu^2 + 3}{24} \left( r_+ + r_- - \frac{\sqrt{r_+ r_- (3 + \nu^2)}}{\nu} \right)
\ee
we get \refe{changecoords}.

\section*{Appendix 2: Modular invariance}\label{modular}
 \addcontentsline{toc}{section}{Appendix 2: Modular invariance}
We present the details of the modular invariance property of BTZ and WBTZ black holes.
 For this section, we are working in Euclidean signature. The Wick rotation in TMG is given by
  \begin{equation}\label{WickrotationTMG}
  t\rightarrow i t_E \quad ,\quad r_-\rightarrow -i r_-^E \quad , \quad \mu\rightarrow - i\mu_E. 
  \end{equation}

The BTZ coordinates have the following periodicities \cite{Carlip:1994gc}
   \begin{equation}\label{periodicityBTZ}
  (t,\phi)\sim(t+\beta,\phi+\Phi)\sim (t,\phi+2\pi)
    \end{equation}
   with 
   \begin{equation}
   \beta= \frac{2 \pi  l^2 r_+}{(r_-^E)^2+r_+^2} \quad , \quad \Phi=-\frac{2 \pi  l r_-^E}{(r_-^E)^2+r_+^2}.
   \end{equation} 
  The first one is needed to avoid a conical singularity at the event horizon and the second is the required identifications to have a black hole rather than a black string. It is useful to repackage the thermal and angular potential in a complex parameter $\tau$, called modular parameter
    \begin{equation}\label{tau}
    \tau\equiv\frac{1}{2\pi}(\Phi+ i \frac\beta l).
    \end{equation}
For BTZ, it is 
    \begin{equation}
    \tau_{BTZ}=-\frac l{r_-^E+i rp}.
    \end{equation}
   
The BTZ black hole is locally AdS, so we can write a change of coordinates to bring the BTZ metric to the one of AdS.
Performing the following change of coordinates\cite{Kraus:2006wn} 
\begin{equation}
      \label{AdSBTZchange}
           \tilde t= \frac{r_{-}Ê  t}{l}+r_+ \phi \quad,\quad
           \rho=l \sqrt{\frac{r^2-r_+^2}{(r_-^E)^2+r_+^2}} \quad,\quad
            \tilde \phi=\frac{r_-^E \phi }{l}-\frac{r_+ t}{l^2},
      \end{equation}
the resulting metric becomes that of euclidean AdS with certain identifications on $(\tilde t , \tilde \phi)$.      
 The first identification of \eqref{periodicityBTZ} implies that
    \[ ( \tilde t,\tilde \phi) \sim 
    ( \tilde t, \tilde \phi -2\pi).\]
while from the second yields
     \[ ( \tilde t,\tilde \phi) \sim 
     ( \tilde t +\beta_{AdS}, \tilde \phi +\Phi_{AdS})
     \]
     with 
     \begin{equation}\label{AdSperiodicities}
 \beta_{AdS}=2\pi r_+ \quad,\quad \Phi_{AdS}=2\pi \frac{r_-^E} l .    
     \end{equation}
We therefore end up with a thermal rotating AdS at temperature and angular velocity \eqref{AdSperiodicities}.
The corresponding modular parameter \eqref{tau} is 
    \begin{equation}
    \tau_{AdS}= \frac{r_-^E + i r_+}l,
    \end{equation}
   related to the one of BTZ by 
   \be
   \tau_{BTZ}=-1/\tau_{AdS}.
   \ee
In short, we have rederived the equivalence between a BTZ black hole at $\tau$ and a thermal rotating AdS space-time at $-1/\tau$\cite{Kraus:2006wn}:
\begin{equation}
ds^2[BTZ(\tau)]=ds^2\left [AdS\left (-\frac1\tau\right )\right ].
\end{equation}
    
We now discuss the WBTZ case of section \ref{wbtztmg}. We saw that the thermodynamic potentials are the same as BTZ.
Moreover, the change of variables \eqref{AdSBTZchange} brings the WBTZ metric to the WAdS timelike one.
As we have the same modular parameter and the same change of variables, we have the same modular invariance property. In other words, WBTZ black hole at a particular $\tau$ is equivalent to the thermal rotating background WAdS at $-1/\tau$, 
 \begin{equation}
 ds^2[WBTZ(\tau)]=ds^2\left [WAdS\left (-\frac1\tau\right )\right ].
 \end{equation}

\end{document}